\documentclass{article}
\usepackage{amsthm}
\usepackage{xcolor}
\usepackage{amsmath}

\usepackage{amsmath, amsfonts}
\usepackage{subfig}
\usepackage{hyperref}
\usepackage{graphicx}
\usepackage{placeins}
\newcommand{\ubeta}{\overline{\beta}}
\newcommand{\ugamma}{\overline{\gamma}}

\DeclareMathOperator*{\argmin}{arg\,min}
\newtheorem{definition}{Definition}
\newtheorem{theorem}{Theorem}
\newtheorem{lemma}{Lemma}
\title{\LARGE \bf Capacity Allocation and Pricing of High Occupancy Toll Lane Systems with Heterogeneous Travelers
}

\author{Haripriya Pulyassary*, Ruifan Yang*, Zhanhao Zhang*, Manxi Wu
\thanks{Authors with * have equal contributions, and are ordered alphabetically. All authors are affiliated with the School of Operations Research and Information Engineering at Cornell University. Email contacts: \{hp297, ry298, zz564, mw843\}@cornell.edu}
}

\begin{document}

\maketitle
\thispagestyle{empty}
\pagestyle{empty}

\begin{abstract}
In this article, we study the optimal design of High Occupancy Toll (HOT) lanes. In our setup, the traffic authority determines the road capacity allocation between HOT lanes and ordinary lanes, as well as the toll price charged for travelers who use the HOT lanes but do not meet the high-occupancy eligibility criteria. We build a game-theoretic model to analyze the decisions made by travelers with heterogeneous values of time and carpool disutilities, who choose between paying or forming carpools to take the HOT lanes, or taking the ordinary lanes. Travelers' payoffs depend on the congestion cost of the lane that they take, the payment and the carpool disutilities. We provide a complete characterization of travelers' equilibrium strategies and resulting travel times for any capacity allocation and toll price. We also calibrate our model on the California Interstate highway 880 and compute the optimal capacity allocation and toll design.


\end{abstract}

\section{INTRODUCTION}
High Occupancy Toll (HOT) lanes are traffic lanes or roadways that are open to vehicles satisfying a minimum occupancy requirement, but also offer access to other vehicles with a toll price. In practice, HOT lanes have been implemented on several interstate highways in California, Texas, and Washington states. With the proper design of lane capacity and toll price, HOT lanes can effectively mitigate traffic congestion through incentivizing carpooling and transit use, while also generate revenue to support transportation infrastructure through toll collection.

The goal of our work is to study the optimal design of HOT lane systems and its impact on traffic congestion. In our model, a traffic authority designs the HOT lane systems by choosing the road capacity of HOT lanes, and the toll price. Given the design of HOT, we develop a game-theoretic model to analyze the strategic decisions made by travelers who have the action set of paying or carpooling to use the 
HOT lane, or using
the ordinary lane. Travelers are modeled as a population of nonatomic players with a continuous distribution of value of time and carpool disutility. Both the HOT lanes and the ordinary lanes are congestible in that the travel time of each lane increases with the aggregate flow induced by travelers' decisions. The outcomes of the system in terms of average travel time cost and toll collection are jointly determined by travelers' equilibrium strategies and the design by the traffic authority . 

We provide a complete characterization of Wardrop equilibrium in this game. In particular, we identify two qualitatively distinct equilibrium regimes that depend on the traffic authority's design of lane capacity and toll price. In the first equilibrium regime, all travelers who take the HOT lane form carpools and no one pays the toll due to the relatively high toll price. In the second equilibrium regime, a fraction of travelers with high carpool disutilities and high value of times make toll payment to take the HOT lanes, while the rest either form carpools or take the ordinary lanes. In both regimes, travelers are split between taking the HOT lanes and the ordinary lanes. 

The equilibrium characterization provides the system designer with insights on how the equilibrium flows and travel time costs of both the HOT lanes and the ordinary lanes depend on the system parameters that include travel time cost functions, capacity allocation and toll price. Moreover, we present comparative static analysis on how the equilibrium flow and costs change with the fraction of capacity that is allocated to the HOT lanes.


We calibrate our model using the data collected from the express lanes of the California Interstate highway 880. We compute the equilibrium strategy profile for a set of discretized design parameters of capacity allocation and toll price. We also compute the Pareto front of the design of HOT lanes, which demonstrates the system authority's tradeoff between minimizing road congestion and maximizing the total toll revenue. 

\emph{Related Literature}. Our model and analysis build on the rich literature of congestion games that includes the equilibrium analysis of routing strategies made by atomic agents \cite{monderer1996potential, rosenthal1973class} and nonatomic agents \cite{sandholm2001potential} in networks, and the analysis on the price of anarchy \cite{roughgarden2004bounding, correa2004selfish, roughgarden2005selfish}. Most of the classical results in congestion games have focused on the settings where all agents have homogeneous preferences. The papers \cite{milchtaich1996congestion, mavronicolas2007congestion} extended these results to study the equilibrium existence and efficiency with player-specific costs.

Previous literature has studied the problem of optimal design of tolling mechanism that minimizes the total travel time cost of nonatomic agents who have homogeneous preferences in both static and dynamic settings \cite{como2021distributed, poveda2017class, paccagnan2019incentivizing, maheshwari2022dynamic}. The papers \cite{fleischer2004tolls, coletolls} have extended the optimal toll design to incorporates travelers' heterogeneous value of time. Moreover, the papers \cite{acemoglu2004flow, correa2018network} have studied tolling and resulting equilibrium costs when tolls on edges are set by strategic operations whose goal is to maximize the toll revenue.

Our paper extends the literature of routing games and toll design to incorporate carpooling and travelers with both heterogeneous value of times and heterogeneous carpool disutilities. Moreover, our work also contributes to the previous studies on the design of HOT lanes that incorporates travelers' choice mode \cite{YANG}, and heterogeneous carpool disutilities \cite{konishi2010carpooling}. However, the models and analysis of these works do not incorporate both the congestion effect, and the strategic behavior of traveler with heterogeneous value of time and carpool disutilities.




\section{Model}

Consider a model of a single, multi-lane highway segment consisting of \emph{ordinary} and \emph{high occupancy toll} lanes. An ordinary lane is toll-free and open to all vehicles. A high occupancy toll lane is accessible without a payment for vehicles that meet a pre-specified minimum occupancy requirement, denoted as $A \geq 2$, but also allow for other vehicles to use with a toll payment. A transportation authority determines the toll price $\tau>0$, the minimum occupancy requirement $A$, and the fraction of road capacity $\rho \in [0,1]$ that is allocated to the HOT lanes. The remaining $(1 - \rho)$-fraction of the total capacity is allocated to the ordinary lanes. 

 
We model the set of travelers as a population of non-atomic agents with total demand of $D>0$. Agents have heterogeneous value of time, denoted as $\beta$, and carpool disutility, denoted as $\gamma$, which is a fixed cost incurred when taking carpool trips. We assume that agents' preference parameters $(\beta, \gamma)$ are uniformly distributed over the set ${ \mathcal{P} := [0, \ubeta] \times [0, \ugamma]}$ where $\ubeta, \ugamma$ are finite and non-negative numbers that represent the maximum value of time and the maximum carpool disutility, respectively. 

The action set of each agent includes:
\begin{itemize}
    \item[-]$R_{toll}$: Use the HOT lane and pay the toll.
    \item[-]$R_{pool}$: Use the HOT lane in a carpool of size $A$.
    \item[-]$R_o$: Use the ordinary lane.
\end{itemize}

Each agent chooses an action in the set $\{R_{toll}, R_{pool}, R_o\}$ based on their preference parameters. Therefore, a strategy profile of this game is a mapping $s: \mathcal{P} \to \{R_{toll}, R_{pool}, R_o\}$. Given $s$, the induced outcome of the game can be represented as a vector $(\sigma_{toll}, \sigma_{pool}, \sigma_{o})$, where $\sigma_{toll}$ (resp. $\sigma_{pool}$, $\sigma_{o}$) is the fraction of the population that plays $R_{toll}$ (resp. $R_{pool}$, $R_o$). Here, $\sigma$ must satisfy $\sigma_{toll}, \sigma_{pool}, \sigma_{o} \in [0, 1]$ and \(\sigma_{toll} + \sigma_{pool} + \sigma_{o} = 1\). Given $\sigma = (\sigma_{toll}, \sigma_{pool}, \sigma_{o})$, we can write the aggregate flow of vehicles that take the ordinary lane, denoted as $x_o$, and the HOT lane, denoted as $x_h$, as follows: 
\begin{align*}
x_o = \sigma_{o} D, \quad x_h=\left(\sigma_{toll} + \frac{\sigma_{pool}}{A}\right)D.
\end{align*}

Given the capacity allocation $\rho$ and the aggregate flow $x=(x_o, x_h)$, the latency function of the HOT lanes is $C_h(x_h; \rho)$, and the latency function of the ordinary lanes is $C_o(x_o; 1 - \rho)$, respectively. We make the following assumptions on the latency functions: 
\begin{enumerate}
\item[(A1)] The latency function $C_o(x_o; 1 - \rho)$ (resp. $C_h(x_h; \rho)$) is increasing in $x_o$ (resp. $x_h$), and decreasing (resp. increasing) in $\rho$. 
\item[(A2)] $C_o(0, 1 - \rho) = C_h(0, \rho)$ for any $\rho \in [0, 1]$.
\end{enumerate}

\medskip

Assumption (A1) shows that the latency of each lane increases in the aggregate flow on that lane, and decreases in the lane capacity.\footnote{This assumption is supported by queueing based models \cite{lu2018probabilistic,lu2022analytical} and empirical validations \cite{national1950highway}.} Assumption (A2) shows that both lanes have identical \emph{free-flow travel time}, which is defined as the value of the latency function with zero flow. 

Given any $\sigma$, we denote the cost of an agent with preference parameters $(\beta, \gamma)$ for playing action $R_i \in \{R_{toll}, R_{pool}, R_o\}$ as $c^\beta_\gamma(R_i; \sigma)$. Then,
\begin{align*}
    c^\beta_\gamma(R_{toll}; \sigma) &= \beta \cdot C_h\left(\left(\sigma_{toll} + \frac{\sigma_{pool}}{A}\right)D; \rho\right) + \tau,\\
    c^\beta_\gamma(R_{pool}; \sigma) &= \beta \cdot C_h\left( \left(\sigma_{toll} + \frac{\sigma_{pool}}{A}\right)D; \rho\right) + \gamma,\\
    c^\beta_\gamma(R_o; \sigma) &= \beta \cdot C_o(\sigma_{o} D; 1 - \rho).
\end{align*}


\begin{definition} A strategy profile $s^*: \mathcal{P} \rightarrow \{R_{toll}, R_{pool}, R_o\}$ is a Wardrop equilibrium if 
\[ s^*(\beta, \gamma) = R_i ~ \Rightarrow ~ c^\beta_\gamma(R_i; \sigma^*) = \argmin_{j \in \{toll, pool, o\}} c^\beta_\gamma(R_j; \sigma^*),\]
where 
\begin{equation}
    D \sigma^*_i = \iint_{\{ (\beta, \gamma) : s^*(\beta, \gamma) = R_i\}} \frac{1}{\bar{\beta}\bar{\gamma}} d\beta d\gamma.   \label{eq: equilibrium condition}
\end{equation}
\end{definition}

\medskip

\section{Equilibrium characterization}

In this section, we characterize the Wardrop equilibrium of the game. For ease of exposition, for any $\sigma$, we define
\[C_\delta(\sigma; \rho) := C_o(\sigma_{o} D; 1 - \rho) - C_h((\sigma_{toll} + \frac{\sigma_{pool}}{A})D; \rho),\] abbreviated as $C_\delta$ when $\sigma$ is clear from the context. This is the difference between the latency of the original lane, and the latency of the HOT lane given $\sigma$.  
\medskip 

In the next lemma, we characterize agents' best response strategies.
\begin{lemma}\label{lem:BestResponse}
For a given $\sigma$, define $\Lambda_i(\sigma) \subseteq \mathcal{P}$ to be the subset of agents whose best response to $\sigma$ is $R_i$, for $i \in \{toll, pool, o\}$. Then,
\begin{align*}
    \Lambda_{toll}(\sigma) &= \{(\beta, \gamma): \beta C_\delta(\sigma; \rho) \geq \tau, ~ \gamma \geq \tau \} \\
     \Lambda_{pool}(\sigma) &= \{(\beta, \gamma): \beta  C_\delta(\sigma; \rho) \geq \gamma, ~ \gamma \leq \tau \} \\
     \Lambda_o(\sigma) &= \{(\beta, \gamma): \beta  C_\delta(\sigma; \rho)\leq \min\{\tau, \gamma\} \}\\
\end{align*}
\end{lemma}

\begin{figure}[ht]
    \centering
    \includegraphics[width=.7\linewidth]{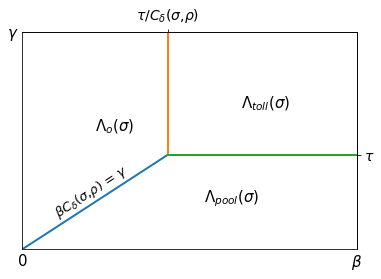}
    \caption{Characterization of best response strategies}\label{fig:decisionBoundary}
\end{figure}


Figure \ref{fig:decisionBoundary} illustrates $\Lambda_{toll}(\sigma), \Lambda_{pool}(\sigma), \Lambda_o(\sigma)$ for an outcome $\sigma$. An agent $j \in \Lambda_i(\sigma)$ if and only if its type $(\beta, \gamma)$ lies in region $\Lambda_i$ in Figure \ref{fig:decisionBoundary}. Notice that, if an agent's  value for time and carpool disutility are both relatively high, the agent prefers the HOT lane over the ordinary lane (due to their high value for time) and prefers paying a toll $\tau$ to carpooling (due to their high carpool disutility) -- so their best response is to play $R_{toll}$. Similarly, if an agent's value for time is high but carpool disutility is at most $\tau$, their best response is to carpool to take the HOT lane (i.e. play $R_{pool}$). If an agent's value of time is relatively low, they do not have incentive to pay $\tau$ to use the HOT lane; if their carpool disutility is low, they choose $R_{pool}$,  otherwise, their choose to take the ordinary lane (i.e. play $R_o$). Finally, since agents' preference parameters are uniformly distributed, $\sigma^*_{toll}, \sigma^*_{pool}, $ and $\sigma^*_{o}$ are equal to the size of $\Lambda_{toll}, \Lambda_{pool}, $ and $\Lambda_o$, respectively.


Due to assumptions (A1) and (A2) on the latency functions $C_o$ and $C_h$, 
both lanes are used in equilibrium, and some (if not all) HOT-users carpool.
\smallskip
\begin{lemma}\label{lem:possibleOutcomes}
$\sigma^*_{pool} > 0$, $\sigma^*_{o} > 0$. 
\end{lemma} 
\smallskip
\begin{proof} 
We show that a strategy profile where all travelers exclusively use a single lane is not an equilibrium. If all flow is routed exclusively on the ordinary lane, agents with low carpool disutility (e.g. $\gamma = 0$) will have the incentive to deviate to the HOT lane, as (A1) and (A2) imply that the latter has strictly smaller latency (as $ \beta C_o(1; 1 - \rho) > \beta C_o(0; 1 - \rho) = \beta C_h(0; \rho)$). The case when traffic is routed exclusively on the HOT lane is symmetric; any agent has an incentive to deviate as the ordinary lane has strictly smaller latency ($ \beta C_h(1;  \rho) + \gamma > C_h(0; \rho) =  \beta C_o(0; 1 - \rho)$). Hence, an equilibrium necessarily routes flow on both lanes, i.e., $\sigma^*_{o} > 0$ and $\sigma^*_{toll} + \sigma^*_{pool} > 0$. 


It remains to show that $\sigma^*_{pool} > 0$. Suppose, to arrive at a contradiction, that $\sigma^*_{pool} = 0$. For any agent with carpool disutility less than $\tau$, $R_{toll}$ is strictly dominated by $R_{pool}$, and thus every such agent must play $R_o$. But for $R_o$ to be a best response,   $\beta C_0(\sigma^*_{o}; 1 - \rho) \leq  \beta C_h(\sigma^*_{toll}; \rho) + \gamma $ for all $\gamma \in (0, \tau)$. Consequently, there must exist some $ \gamma^\dagger$ such that for all $\beta \in [0, \bar \beta]$, $\beta C_0(\sigma^*_{o}; 1 - \rho) <  \beta C_h(\sigma^*_{toll}; \rho) + \gamma^\dagger \leq  \beta C_h(\sigma^*_{toll}; \rho) + \tau $; that is, any agent playing $R_{toll}$ has incentive to deviate to $R_o$. Hence, $\sigma^*_{toll} = 0$. This contradicts the above assertion that $\sigma^*_{toll} + \sigma^*_{pool} > 0$. 
\end{proof}
\medskip

Lemma \ref{lem:possibleOutcomes} ensures that in equilibrium both lanes are used, and either (A) all HOT-users meet the minimum occupancy requirement, or (B) some (but not all) users of the HOT-lane pay the toll $\tau$. This suggests the following two qualitatively distinct equilibrium regimes.
\begin{enumerate}
    \item \emph{Regime A}: The toll price $\tau$ is relatively high. 
    All agents who take HOT lane carpool. \[
            \left\{ (\rho, \tau): \tau > \min\left\{\ugamma,~ \ubeta C_\delta\left(0, \frac{\tau}{2\bar \gamma}, 1 -  \frac{\tau}{2\bar \gamma}, \rho\right)\right\} \right\}.                                 
        \]
    \item \emph{Regime $B$}: The toll price $\tau$ is relatively low. Some agents pay $\tau$ to use the HOT lane. 
        \[ \left\{ (\rho, \tau) : \tau < \min\left\{\ugamma, ~ \ubeta  C_\delta\left(0, \frac{\tau}{2\bar \gamma}, 1 -  \frac{\tau}{2\bar \gamma},  \rho\right)\right\} \right\}. \]
\end{enumerate}

 \begin{theorem} \label{thm:NE}
The Wardrop equilibrium is unique in each regime and can be written as follows:

\noindent \underline{Regime $A$}: $\sigma_{toll}^*=0$.\\ 
    \begin{enumerate}
        \item If $\ugamma > \ubeta C_\delta(0, \frac{\tau}{2\ugamma}, 1 - \frac{\tau}{2\ugamma}; \rho) $, then $\sigma^*_{pool}$ is solved by
        \begin{equation} \label{eq:b2-fixed_point1}
            \frac{1}{2} \frac{\ubeta}{\ugamma} C_\delta(0, \sigma^*_{pool}, 1 - \sigma^*_{pool}; \rho) = \sigma^*_{pool},
        \end{equation}
        and $\sigma^*_{o} = 1 - \sigma^*_{pool}$.
        \item  If $\ugamma < \ubeta C_\delta(0, \frac{\tau}{2\ugamma}, 1 - \frac{\tau}{2\ugamma}; \rho)$, then $\sigma^*_{o}$ is solved by
        \begin{equation} \label{eq:b2-fixed_point2}
            \frac{1}{2} \frac{\ugamma}{\ubeta} \frac{1}{C_\delta(0, 1-\sigma^*_{o}, \sigma^*_{o}; \rho)} = \sigma^*_{o},
        \end{equation}
         and $\sigma^*_{pool} = 1 - \sigma^*_{o}$.
    \end{enumerate}

\noindent \underline{Regime $B$}: 
        $\sigma^*_{toll}$ is solved by
        {\begin{equation}
        \left ( 1 - \frac{\tau}{\ubeta C_{\delta}\left(\sigma^*_{toll}, \sigma^*_{pool}, \sigma^*_{o}; \rho\right )}\right) \cdot (\frac{\ugamma - \tau}{\ugamma}) = \sigma^*_{toll}\label{eq:b3-fixed_point},
        \end{equation}}
        and 
        \begin{align*} 
            \sigma^*_{pool} = \frac{\tau}{2} (\frac{1}{\ugamma - \tau}\sigma^*_{toll} + \frac{1}{\ugamma}), && 
	    \sigma^*_{o} = 1 - (\sigma^*_{toll} + \sigma^*_{pool}).
        \end{align*}

\end{theorem}

\medskip 


%


\begin{figure}[ht]
    \centering    
    \subfloat[\centering $\tau > \ubeta C_\delta\left(0, \frac{\tau}{2\bar \gamma}, 1 -  \frac{\tau}{2\bar \gamma}, \rho\right)$]{{\includegraphics[width=0.47\linewidth]{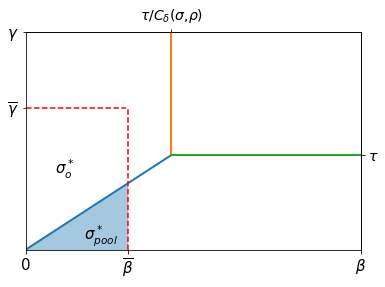}} \label{fig:A1}}
    \subfloat[\centering $\tau > \ugamma$]{{\includegraphics[width=0.47\linewidth]{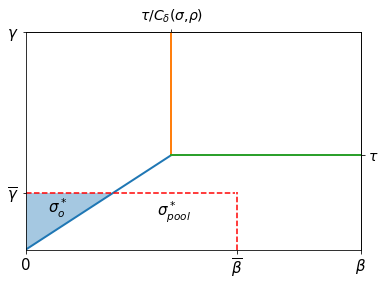}}\label{fig:A2} }
    \caption{Equilibrium outcome in regime A}
\end{figure}


In regime $A$, there are no single-occupancy vehicles on the HOT lane, that is, all agents choose $R_{pool}$ or $R_o$.  In this regime, either $\ubeta$ is small enough such that for any agent the cost of carpooling is smaller than the cost of paying the HOT price (case A1 as in Figure \ref{fig:A1}), or $\ugamma$ is small enough such that for any agent the cost of the ordinary lane is smaller than the cost of paying the HOT price (case A2 as in Figure \ref{fig:A2}).

In case $A1$, $\sigma^*_{pool}$ is solved by\eqref{eq:b2-fixed_point1} since $\sigma^*_{pool}$  must equal to fraction of the size of the shaded triangle in Figure \ref{fig:A1} that represents the set of agents who choose carpool over the entire set, which is represented by the dotted rectangle. The remaining agents choose the ordinary lane. Similarly, in case $A2$, $\sigma^*_{o}$ is solved by \eqref{eq:b2-fixed_point2} since $\sigma^*_{o}$ must equal to fraction of the size of the shaded triangle in Figure \ref{fig:A2} that represents the set of agents who choose ordinary over the entire dotted rectangle. The remaining agents choose to take carpool.

\color{black}

  
\begin{figure}[ht]
    \centering
    \includegraphics[width=0.7\linewidth]{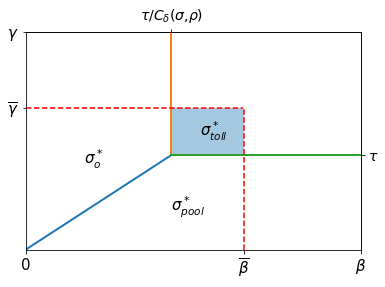}
    \caption{Equilibrium outcome in regime B}
    \label{fig:B}
\end{figure}

In Regime B, all three actions are chosen by players. As illustrated in Figure \ref{fig:B}, the entire set (dotted red region) overlaps with all three best response regions. Equation \eqref{eq:b3-fixed_point} represents that the ratio between the size of the shaded rectangular area and the entire set is the fraction of agents who pay tolls to take the HOT lanes, i.e. the value of $\sigma_{toll}$. 
The shaded square has base $\ubeta - \tau / C_{\delta}\left(\sigma^*; \rho\right )$ and height $\ugamma - \tau$.

Next, we analyze how the Wardrop equilibrium and the size of each regime change with the fraction of capacities allocated to the HOT lane $\rho$. 


\begin{theorem}\label{theorem:comparative}
    Consider a Wardrop equilibrium with design parameters $(\rho, \tau)$. Let $(\sigma^*_{toll},\sigma^*_{pool},\sigma^*_{o})$ be the corresponding demands for each strategy at equilibrium. For any fix $\tau$, as we increase $\rho$, 
    \begin{enumerate}
        \item The difference in latency between the ordinary lane and the HOT lane $C_\delta(\sigma^*, \rho)$ increases.
        \item The size of regime A is non-increasing and the size of regime B is non-decreasing.
        \item $\sigma^*_{toll}$ is non-decreasing, $\sigma^*_{pool}$ is decreasing, and $\sigma^*_{o}$ is increasing.
    \end{enumerate}
\end{theorem}
Theorem \ref{theorem:comparative} is intuitive -- as we increase the capacity of HOT lane, more people are incentivized to use the HOT lane, and fewer are incentivized to use the original lane. As a result, the difference between the latencies of the two lanes decrease. 


\section{Optimal design of HOT on California I-880}
\begin{figure}[htp]
    \centering
    \includegraphics[width=0.7\linewidth]{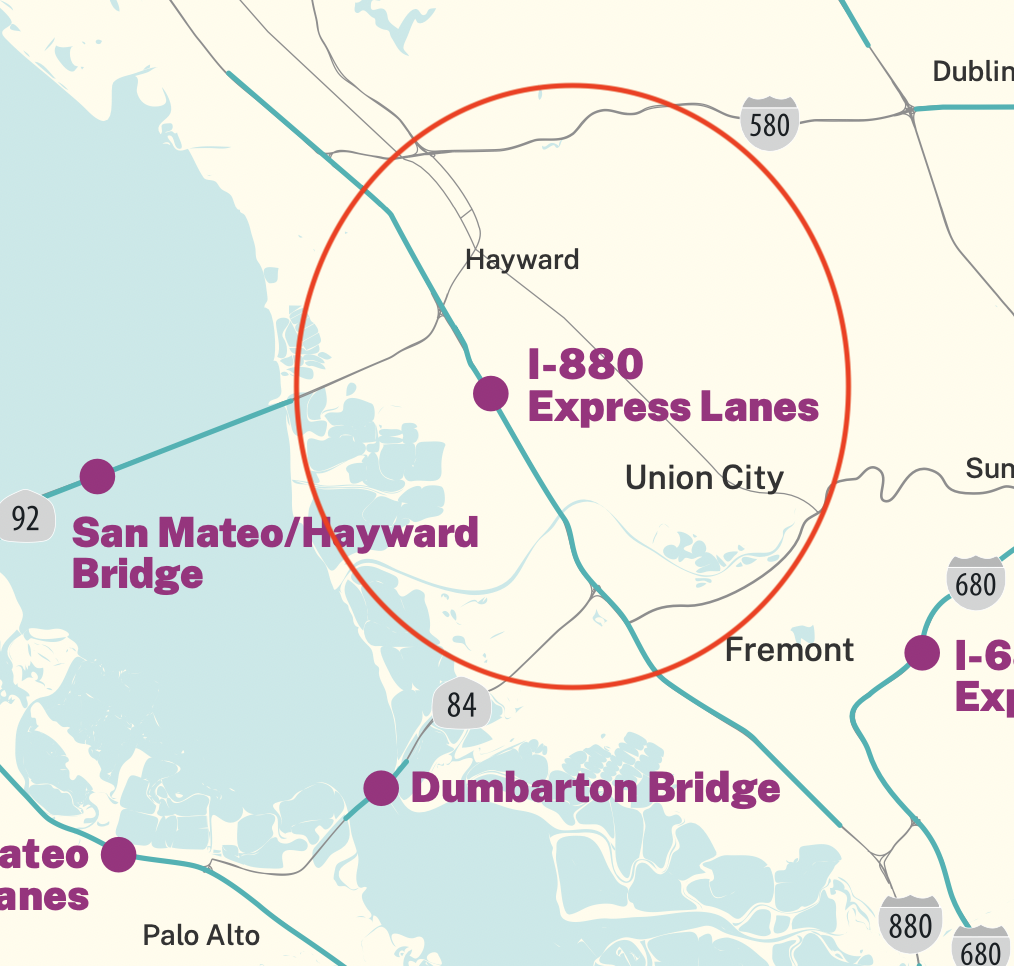}
    \caption{Interstate 880 (I-880) Highway (Encircled in Red)}
    \label{fig:map}
\end{figure}

We calibrate our equilibrium analysis using the data collected from the Northbound of Interstate 880 Highway (I-880) (Fig. \ref{fig:map}) between Dixon Landing Road and Lewelling Blvd. A fraction of the segment has minimum occupancy requirement of 2 and the remaining segment has minimum occupancy requirement of 3. In this case study, we take $A=2.5$.


We adopt the Bureau of Public Roads (BPR) function to characterize the average travel cost per driver (free flow travel time plus congestion time) \cite{ccolak2016understanding}. Given the HOV capacity $\rho$ and an outcome $\sigma = (\sigma_{toll}, \sigma_{pool}, \sigma_o)$, the latency function is given by the BPR function as follows:
\begin{align*}
    &C_o(\sigma_o, \rho) = T_f \cdot \left[ 1 + \left( a \cdot \frac{D \cdot \sigma_o}{V \cdot (1 - \rho)}\right)^b\right]\\
    &C_h(\sigma_{toll}, \sigma_{pool}, \rho) = T_f \cdot \left[ 1 + \left(a \cdot \frac{D \cdot (\sigma_{toll} + \sigma_{pool} / A)}{V \cdot \rho}\right)^b\right]
\end{align*}
where $a$ and $b$ are the BPR coefficients, $T_f$ is the free flow travel time, and $V$ is the capacity. We set $a = 0.15$ and $b = 4.0$ following \cite{roads1964traffic, national1950highway}. From the traffic flow data collected by the California Department of Transportation \footnote{https://pems.dot.ca.gov/}, the avergae traffic demand is $D = 115$ vehicles per minute. Since the highway segment has four lanes, we estimate the capacity as $V = 140$ vehicles per minute following \cite{ccolak2016understanding}. We use Google Maps to estimate the free-flow travel time as $T_f = 22$ minutes. 

We set the maximum carpool disutility as $\ugamma = \$8$. This is estimated by taking the difference between the average trip price of Uber with single customer and the average price of a carpool trip. We estimate the maximum value of time as $\ubeta = \$1.5/\text{min}$ following the analysis of income-based Bay area value of time calibration in \cite{wu2023}. 

We measure the total congestion by the average travel time of all travelers: \[
    T(\tau, \rho) := (\sigma_{toll} + \sigma_{pool}) \cdot C_{h}(\sigma_{toll}, \sigma_{pool}, \rho) + \sigma_o \cdot C_o(\sigma_o, \rho).
\]

Additionally, we can compute the total toll revenue as \[
    R(\tau, \rho) := D \cdot \sigma_{toll} \cdot \tau.
\]


We discretize the feasible region of toll price $\tau$ and the capacity fraction $\rho$. In particular, we choose the set $\rho \in\{ \frac{1}{4}, \frac{2}{4}, \frac{3}{4}\}$ since the highway has four lanes. Additionally, we set the toll price within the range of $0.5$ and $10$ dollars discretized by $\$0.5$. 

\begin{figure}[htp]
    \centering
    \includegraphics[width=\linewidth]{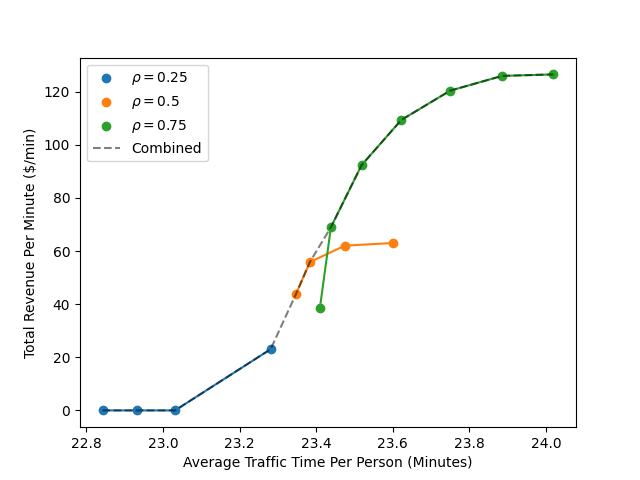}
    \caption{Pareto Front of $T(\tau, \rho)$ and $R(\tau, \rho)$.}
    \label{fig:pareto_sep}
\end{figure}

For each pair of $(\tau, \rho)$, we compute the equilibrium strategies, the average driving time $T(\tau, \rho)$ and the total toll revenue $R(\tau, \rho)$. In Figure \ref{fig:pareto_sep}, we plot the Pareto front of the tuple of $(T, R)$ (marked as the black dashed line), which demonstrates the tradeoff between congestion minimization and toll revenue faced by the authority. Additionally, we plot the Pareto front of $(T, R)$ for each feasible capacity allocation $\rho$. We can see that as we increase the number of HOV lanes, we effectively increase the toll revenue with cost of increasing the average travel time of all drivers.

\section{Concluding remarks}
In this article, we examine a game-theoretic model that analyzes the lane choice oftravelers with heterogeneous values of time and carpool disutilites on highways equipped with HOT lanes. We characterize the equilibrium strategies, and identify two qualitatively distinct equilibrium regimes that depends on the HOT lane capacity and toll price. We calibrate our model using the data of California Interstate highway 880 and determined the optimal capacity allocation and toll design. As a future direction of research, we will extend our analysis to non-uniform distribution of preference parameters, and to incorporate traffic flows with multiple origins and destinations.

\bibliographystyle{unsrt}
\bibliography{references}


\end{document}